\documentclass[aps,pre,showpacs,twocolumn]{revtex4}
\usepackage{amssymb}
\usepackage{graphicx}
\usepackage{colordvi}

\begin{document}

\title{Local boundary conditions for NMR-relaxation in digitized porous media}

\author{M \"{O}gren$^{1,2}$}

\date{\today{}}

\affiliation{$^{1}$Nano Science Center, Department of Chemistry, University of
Copenhagen, Universitetsparken 5, DK-2100 K{\o}benhavn {\O}, Denmark.\\
 $^{2}$School of Science and Technology, \"{O}rebro University, SE-701
82 \"{O}rebro, Sweden.}

\begin{abstract}
We narrow the gap between simulations of nuclear magnetic resonance
dynamics on digital domains (such as CT-images) and measurements in
$D$-dimensional porous media. We point out with two basic domains, the
ball and the cube in $D$ dimensions, that due to a \emph{digital
uncertainty} in representing the real pore surfaces of dimension $D-1$,
there is a systematic error in simulated dynamics. We then reduce
this error by introducing local Robin boundary conditions. 
\end{abstract}

\pacs{02.60.-x, 81.05.Rm, 76.60.-k} 
%{02.60.-x}{Numerical approximation and analysis}
%{81.05.Rm}{Porous materials; granular materials}
%{76.60.-k}{Nuclear magnetic resonance and relaxation}

\maketitle

\section{Introduction}

The dynamics of nuclear magnetic resonance (NMR), pioneered by Felix
Bloch~\cite{Bloch1946} and Henry Torrey~\cite{Torrey1956}, is
an important tool in modern technology, and is now used in hospitals
every day. Apart from magnetic resonance imaging (MRI) in medicine \cite{LinAPL2013}, NMR is also used
to study biofilms in contaminated water \cite{Fridjonsson2011}, and
porous fiber materials in paper, filters and membranes, as well as insulation
materials in industry \cite{Tomadakis2003}. 
For petrophysical applications
\cite{Kenyon1992,Senturia1970,Cohen1982}, such as the early use of
NMR-relaxation times in the famous Kozeny-Carman permeability correlations \cite{Kenyon1992,BanavarPRL1987,Dunn1999},
it is motivated by the interpretation as a surface-to-volume ratio
for the pores of a medium. 
Quantitative agreement between
large scale simulations and measurements of relaxation times is so
far only obtained if the surface relaxation parameter is substantially
\emph{adjusted} \cite{OerenSPE2002,Talabi2009}.
Bergman \emph{et
al.} have pointed out that such problems may arise due to the digital misrepresentation of the true surfaces \cite{BergmannPRE1995},
and we here give a practical solution to this problem for discrete random walks on digital domains.  

The relaxation time is determined by the time for a
directionally excited \emph{magnetic spin} $M\left(\mathbf{x},t\right)$
distribution, carried by the protons of the molecules, to relax towards
equilibrium. 
The differential equation and the Robin boundary condition
(BC) governing the quantitative dynamics under study here are~\cite{Torrey1956,Senturia1970,Cohen1982,GrebenkovRMP2007,BrownsteinPRA1979,SongPRL2008,FinjordTPM2007}
\begin{equation}
\frac{\partial M}{\partial t}=D_{0}\nabla^{2}M-\frac{M}{T_{V}},\: D_{0}\mathbf{n}\cdot\nabla M+\rho M=0.\label{eq:Diffusion_equation}
\end{equation}
Above $D_{0}$ is the diffusion coefficient,
$T_{V}$ is the characteristic time of the volume relaxation, and
$\rho$ is the surface relaxation parameter. Above $T_{V}=T_{V,1}$
or $T_{V}=T_{V,2}$ and $\rho=\rho_{1}$ or $\rho=\rho_{2}$ can describe longitudinal
or transverse components \cite{Kenyon1992,GrebenkovRMP2007}. Realistic
values for the above physical parameters can be found in~\cite{Kenyon1992,Talabi2009}
and references therein. 

Except for $T_{V}$, there are two characteristic
times for a pore of size $R_{0}$, namely the diffusion time, $T_{D_{0}}\sim R_{0}^{2}/D_{0}$,
and the surface relaxation time, $T_{\rho}\sim R_{0}/\rho$, suggesting
two different regimes: \emph{fast diffusion} ($T_{D_{0}}\ll T_{\rho}$)
and \emph{slow diffusion} ($T_{D_{0}}\gg T_{\rho}$). 

The focus here is on \emph{accurate numerical simulations}, not the observables
themselves, and we show results only for the total \emph{magnetization}
$\mathcal{M}\left(t\right)=\int_{\Omega}M\left(\mathbf{x},t\right)d\mathbf{x}$ in this article.
Although NMR correlations \cite{GrebenkovRMP2007}, such as so called $T_{1}-T_{2}$
\cite{SongPRL2008} and $T_{2}-D$ \cite{ArnsNJP2011} correlations, are becoming
important in applications, they suffer from the same need of accurate
large scale simulations. 

We also emphasize
that when $T_{V}\rightarrow\infty$ in Eq.~(\ref{eq:Diffusion_equation}) we
may model many other situations where a chemical reaction or
adsorption takes place at surfaces, for example in crystal dynamics,
where also the boundaries change in time \cite{Hausser2013}. 

With a dimensionless magnetic moment $m\left(\mathbf{x},t\right)=R_{0}^{D}\exp\left(t/T_{V}\right)M\left(\mathbf{x},t\right)/\mathcal{M}\left(0\right)$,
and variables $\xi=r/R_{0}$ and $\tau=D_{0}t/R_{0}^{2}$, we can write Eq.~(\ref{eq:Diffusion_equation}) as 
\begin{equation}
\frac{\partial m}{\partial \tau}=\nabla^{2}m, \: \mathbf{n}\cdot\nabla m+\rho_{0}m=0, 
\label{eq:dimless_Diffusion_equation}
\end{equation}
where $\rho_{0}=R_{0}\rho/D_{0}$ is the dimensionless surface relaxation parameter. 

In the next section we examine diffusion with Robin boundary conditions for a radially symmetric domain with a radial random walk.
We observe perfect agreement with an analytic solution for this case and then later use the digitized circle as a testcase.
In Sec.~\ref{CRW_3} we examine diffusion in a $D$-cube with a Cartesian random walk. 
Again perfect agreement is observed with an analytic solution when the Cartesian lattice have the same orientation as the cube.
To treat general domains, we discuss in section~\ref{LLBC_4} a local boundary correction.
In Sec.~\ref{RFTDD_5} we apply linear local boundary conditions to two-dimensional domains and evaluate the performance of the random walk method numerically.
In the final section we discuss and summarize our results.

\section{A radial random walk for the $D$-ball} \label{RRW_2}

Let us consider the $D$-ball of radius $R_{0}$ with the boundary surface $\partial\Omega$ and domain volume $\Omega$ given by
\begin{equation}
S_{D}\left(R_{0}\right)=\frac{2\pi^{D/2}}{\Gamma\left(D/2\right)}R_{0}^{D-1},\:\: V_{D}\left(R_{0}\right)=S_{D}\left(R_{0}\right)\frac{R_{0}}{D}.\label{eq:SurfaceAndVolumeForDSphere}
\end{equation}
We first model the radial dynamics with a random walk on a discretized radius $\Delta r,\:2\Delta r,\:\ldots,\: R_{0}$.
We choose the probability for a step inwards to be proportional to the relative
decrease in volume for such a step
\begin{equation}
P_{D}\left(r\rightarrow r-\Delta r\right)=\left\{ \begin{array}{l}
\frac{\left(r-\Delta r\right)^{D-1}}{r^{D-1}+\left(r-\Delta r\right)^{D-1}},\: r>\Delta r\\
0,\: r=\Delta r
\end{array}\right.,\label{eq:P_D_for_radial_random_walk}
\end{equation}
and correspondingly for a step outwards, $P_{D}\left(r\rightarrow r+\Delta r\right)=1-P_{D}\left(r\rightarrow r-\Delta r\right)$.
Some authors include also a third probability term $P_{D}\left(r\rightarrow r\right)$
(no radial step) \cite{BoettcherPRL1995}, but it is sufficient to instead
use the one-dimensional diffusion coefficient $D_{0}^{\left(1\right)}=\Delta r^{2}/2\Delta t$ for the radial random walk in any dimension $D$. 
If $r>R_{0}$, the trajectory is annihilated with the surface relaxation
probability $p_{S}=\rho\Delta r/D_{0}^{\left(1\right)}$ (we outline
a related derivation in Sec.~\ref{CRW_3}), otherwise it is reflected. 

\subsection{Uniform initial conditions}
The formalism
presented above does not only provide a numerical scheme, but also the short-time asymptotes for the magnetization of the $D$-ball
with a uniform initial condition. 
The initial number of walkers hitting the
surface is $\Delta N\simeq P_{D}\left(R_{0}\rightarrow R_{0}+\Delta r\right)S_{D}\left(R_{0}\right)\Delta r/V_{D}\left(R_{0}\right)$
which gives with Eq.~(\ref{eq:SurfaceAndVolumeForDSphere}) and the definitions of $D_{0}^{\left(1\right)}$ and $p_S$ that
\begin{equation}
\frac{1}{\mathcal{M}\left(0\right)}\left.\frac{d\mathcal{M}\left(t\right)}{dt}\right|_{t=0}\simeq-p_{S}\frac{\Delta N}{\Delta t}=-\frac{\rho D}{R_{0}}.\label{eq:Asymptote_For_D_Ball}
\end{equation}

To benchmark the role of $\Delta r$ in numerical random walks,
we calculate the analytic solution to Eq.~(\ref{eq:dimless_Diffusion_equation})
for the circle ($D=2$) with a uniform initial condition $m\left(\mathbf{x},0\right)=\pi^{-1}$.
Since Eq.~(\ref{eq:dimless_Diffusion_equation}) is linear, we can apply Sturm-Liouville theory for this radially symmetric problem, and the subsequent expansion of the magnetization is in the two-dimensional case a so called
Dini-series
\begin{equation}
\mathcal{M}\left(t\right)=\mathcal{M}\left(0\right)e^{-t/T_{V}}\sum_{k=1}^{\infty}\frac{4J_{1}^{2}\left(\gamma_{k,0}\right)}{\left(\rho_{0}^{2}+\gamma_{k,0}^{2}\right)J_{0}^{2}\left(\gamma_{k,0}\right)}e^{-\gamma_{k,0}^{2}\tau}.\label{eq:M_uni_Magnetization}
\end{equation}
Here $\gamma_{k,0}$ is the $k$th root of $\xi J_{0}'\left(\xi\right)+\rho_{0}J_{0}\left(\xi\right)=0$,
where $J_{\nu}$ is the first kind Bessel function. 
\begin{figure}
\includegraphics[scale=0.31]{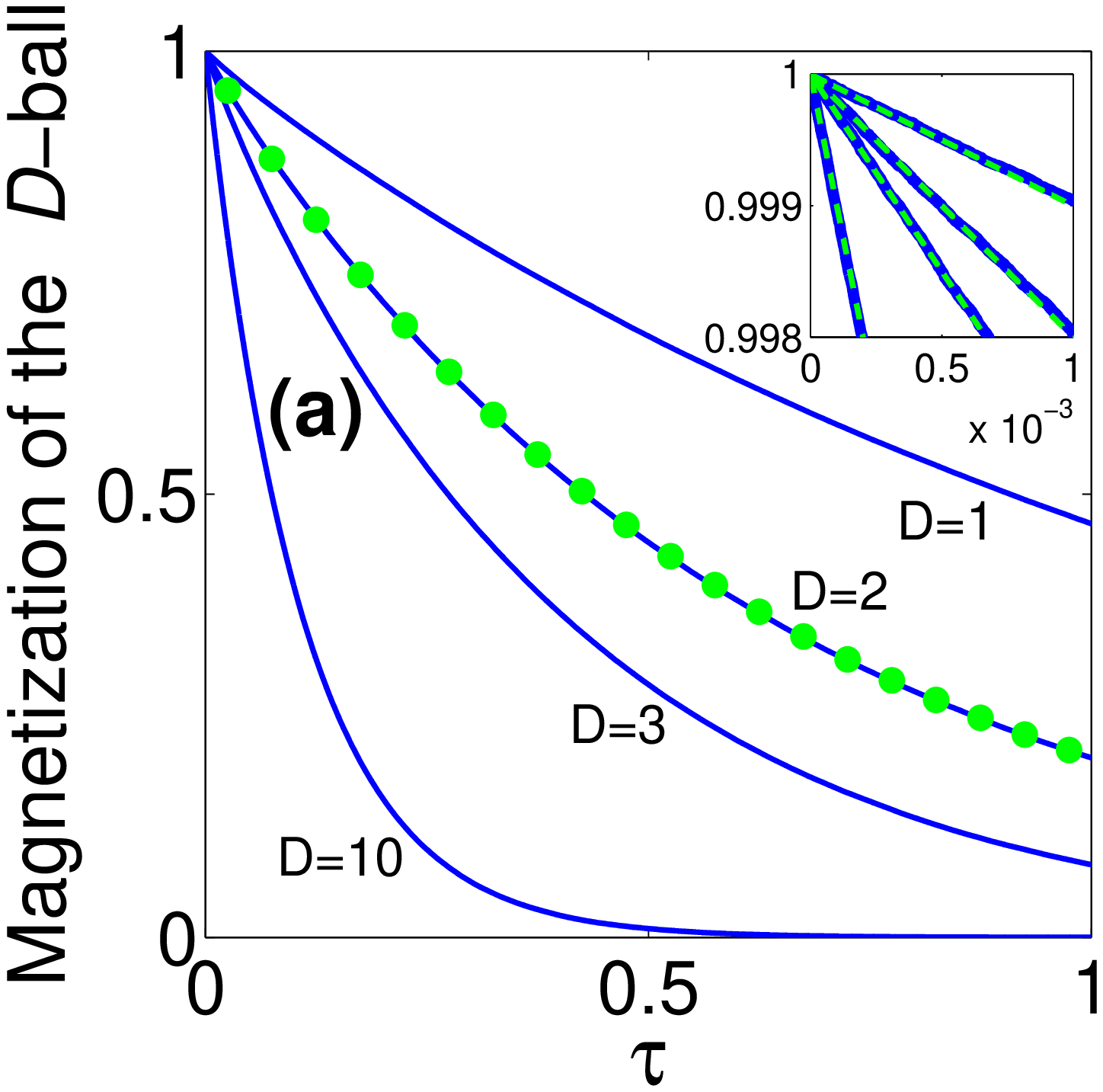}~\hspace{-4mm}~\includegraphics[scale=0.31]{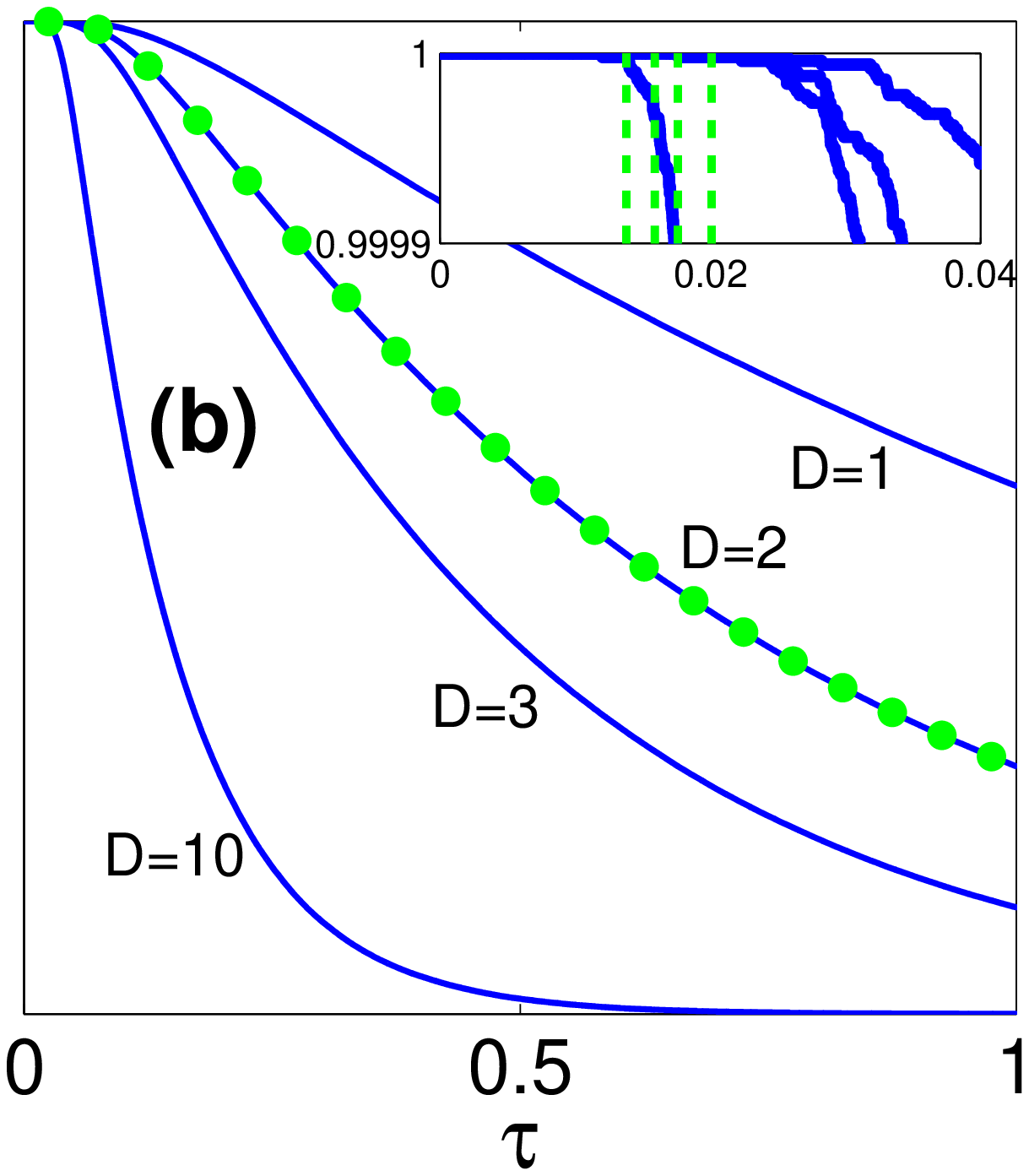}
\caption{(Color online) Dynamics of $\mathcal{M}\left(t\right)$ from Eq.~(\ref{eq:dimless_Diffusion_equation}) integrated over the spatial dimensions for different initial conditions: (a) $m\left(\mathbf{x},0\right)=R_{0}^{D}/V_{D}\left(R_{0}\right)$;
(b) $m\left(\mathbf{x},0\right)=R_{0}^{D}\delta\left(\mathbf{x}\right)$.
Solid (blue) curves shows results from the $D$-dimensional radial
random walk. 
For the two-dimensional case ($D=2$), we also plot the
Dini-series of Eqs.~(\ref{eq:M_uni_Magnetization}) in (a) and (\ref{eq:M_delta_Magnetization}) in (b) with (green) dots.
Inset of (a) shows the short-time asymptotes of Eq.~(\ref{eq:Asymptote_For_D_Ball}),
dashed (green) lines. Inset of (b) shows the first-arrival times $\left\langle t_{1,N}\right\rangle $
(see text) with dashed vertical lines.\label{fig_1}}
\end{figure}

In Fig.~\ref{fig_1} (a) we show the agreement between numerical results from the radial random walk for $D=2$ and the analytic solution of Eq.~(\ref{eq:M_uni_Magnetization}). 
 Physical parameters have been set to unity besides $T_{V}\rightarrow\infty$, which means that we have no volume relaxation.
We used $\Delta r=10^{-2}R_{0}$
and $10^{6}$ initial trajectories throughout where not explicitly stated.

\subsection{Localised initial conditions}
For an initially non-uniform magnetic moment the weight of higher modes
are increasing and there is generally no short-time approximations
available, as in Eq.~(\ref{eq:Asymptote_For_D_Ball}). Motivated by this,
we have in Fig.~\ref{fig_1} (b) benchmarked numerical results also for the circle with
opposite extreme initial conditions, the central delta spike $m\left(\mathbf{x},0\right)=R_{0}^{2}\delta\left(\mathbf{x}\right)$. 
In this case only the numerator in Eq.~(\ref{eq:M_uni_Magnetization}) change, and the total magnetization now reads 
\begin{equation}
\mathcal{M}\left(t\right)=\mathcal{M}\left(0\right)e^{-t/T_{V}}\sum_{k=1}^{\infty}\frac{2\gamma_{k,0}J_{1}\left(\gamma_{k,0}\right)}{\left(\rho_{0}^{2}+\gamma_{k,0}^{2}\right)J_{0}^{2}\left(\gamma_{k,0}\right)}e^{-\gamma_{k,0}^{2}\tau},
\label{eq:M_delta_Magnetization}
\end{equation}
and again agreement is found with the numerical results for $D=2$.

As seen in the inset of Fig.~\ref{fig_1}(b), the initial slope is zero because no walkers are initially close to the
surface. 
The relevant quantity to predict is the time when
the surface relaxation starts.
If $p_{S}\sim1$ it means that practically all trajectories are annihilated at the surface and we have a Dirichlet BC. 
For this case, we have confirmed that the time when
the surface relaxation starts is precisely the first-arrival time $\left\langle t_{1,N}\right\rangle $.
We plot the results of a second-order expansion of $\left\langle t_{1,N}\right\rangle $
\cite{YustePRE2001} in the inset of Fig.~\ref{fig_1}(b), from which
qualitative agreement is observed even though $p_{S}\ll1$ in those numerical examples. 
In comparison,
the $j$th-first-arrivals time $\left\langle t_{j,N}\right\rangle $
\cite{YustePRE2001} for $j\simeq1/p_{S}$ overestimates this time
since early boundary arrivers may bounce close to the surface.

\section{The $D$-cube as a testcase for a Cartesian random walk}  \label{CRW_3}

As the second basic domain, $\Omega$ is chosen to be the $D$-cube, and then it is straightforward to obtain a compact analytic solution for the magnetization valid for any $D$ with Sturm-Liouville theory from Eq.~(\ref{eq:dimless_Diffusion_equation})
\begin{equation}
\mathcal{M}\left(t\right)=\mathcal{M}\left(0\right)e^{-t/T_{V}}2^{D}\prod_{j=1}^{D}\sum_{k_{j}=1}^{\infty}c_{k_{j}}e^{-\lambda_{k_{j}}\tau},\label{eq:Solution_For_D_Cube}
\end{equation}
with coefficients
\begin{equation}
c_{k_{j}}=\left[\frac{\sin\left( \sqrt{\lambda_{k_{j}}} \right) }{\sqrt{\lambda_{k_{j}}}}\right]^{q}\frac{\sin \left( \sqrt{\lambda_{k_{j}}} \right) }{\cos \left( \sqrt{\lambda_{k_{j}}} \right) \sin \left( \sqrt{\lambda_{k_{j}}} \right) +\sqrt{\lambda_{k_{j}}}},
\label{eq:Coefficients_For_D_Cube}
\end{equation}
where $q=1$ ($0$) for uniform (central) initial conditions, while
the eigenvalues $\lambda_{k_{j}}$ fulfill $\sqrt{\lambda_{k_{j}}}\tan\sqrt{\lambda_{k_{j}}}=\rho_{0}$.

We remark that for $D=1$, Eqs.~(\ref{eq:Solution_For_D_Cube}) and
(\ref{eq:Coefficients_For_D_Cube}) agrees with the $D=1$ result
of the radial random walk of Eq.~(\ref{eq:P_D_for_radial_random_walk}) with
the physical domain being $r\in\left[-R_{0},R_{0}\right]$. 

In agreement
with Eq.~(\ref{eq:Asymptote_For_D_Ball}), we can obtain from Eqs.~(\ref{eq:Solution_For_D_Cube})
and (\ref{eq:Coefficients_For_D_Cube}) 
\begin{equation}
\frac{1}{\mathcal{M}\left(0\right)}\left.\frac{d\mathcal{M}\left(t\right)}{dt}\right|_{t=0}= -2^{D}\prod_{j=1}^D\sum_{k_{j}=1}^{\infty}c_{k_{j}}\lambda_{k_{j}}=-\frac{\rho_0 D}{R_{0}},\label{eq:Asymptote_For_D_Cube}
\end{equation}
also for the $D$-cube. 
In fact the asymptotic results of Eqs.~(\ref{eq:Asymptote_For_D_Ball}) and~(\ref{eq:Asymptote_For_D_Cube}),
for the specific geometries presented are generally valid for any
connected pore in $D$ dimensions with a uniform initial condition.
Integrating the dimensionless diffusion equation in Eq.~(\ref{eq:dimless_Diffusion_equation}) over the volume, applying the corresponding Robin BC,
and finally using Gauss's theorem for the divergence we have 
\begin{equation}
\int\frac{\partial m}{\partial\tau}dV=-\rho_{0}\oint mdS =-\frac{\rho_0 S}{V},
\label{eq:Asymptote_General}
\end{equation}
where we have assumed a uniform magnetic moment for all times, i.e., $m\left(\mathbf{x},\tau\right)=m\left(\tau\right)$, in the last step.
The result (\ref{eq:Asymptote_General}) is in agreement with the right hand side of Eqs.~(\ref{eq:Asymptote_For_D_Ball}) and~(\ref{eq:Asymptote_For_D_Cube}) for the $D$-ball and $D$-cube respectively.
Note that for fast diffusion (in relation also to the size and connectedness
of the pores) the magnetic moment is kept approximately uniform and $\mathcal{M}\left(t\right)\sim\exp\left(-\rho St/V\right)$
for any time. Many estimates using the pore-size distribution are
based on this approximation \cite{Kenyon1992,Cohen1982,MendelsonPRB1990}.

We now consider a random walk in a $D$-dimensional
Cartesian lattice for a general porous medium. 
The change in
the fraction of trajectories at a given lattice point during a time step $\Delta t$ is given
by the probability for a step \emph{from} any of the $2D$ neighboring
lattice points, and then the probability for a step \emph{to}
the neighboring points is subtracted. For a point next to a boundary, that has
$n$ neighboring boundary surfaces, we can without loss of generality
for the result assume those to be in the positive Cartesian directions
$x_{j},\: j=D-n+1,...,D$. We consider the change per time $\Delta t$
(with $\Delta r\equiv\Delta x_{1}=...=\Delta x_{D}$ for notational simplicity) 
\begin{widetext}
\begin{displaymath}
%%\label{s.long}
\frac{M\left(\mathbf{x},t+\Delta t\right)-M\left(\mathbf{x},t\right)}{\Delta t}=\frac{\Delta r^{2}}{2D\Delta t}\sum_{j=1}^{D-n}\left[\frac{M\left(\mathbf{x}-\Delta r \mathbf{e}_{j},t\right)-2M\left(\mathbf{x},t\right)+M\left(\mathbf{x}+\Delta r \mathbf{e}_{j},t\right)}{\Delta r^{2}}\right]
\end{displaymath}
\begin{equation}
+\frac{\Delta r}{2D\Delta t}\sum_{j=D-n+1}^{D}\left[\frac{M\left(\mathbf{x}-\Delta r\mathbf{e}_{j},t\right)-M\left(\mathbf{x},t\right)}{\Delta r}\right]-\frac{1}{\Delta t}\frac{np_{S}}{2D}M\left(\mathbf{x},t\right)-\frac{p_{V}\left(\Delta t\right)}{\Delta t}M\left(\mathbf{x},t\right).\label{eq:BC_discrete_derivation_Cartesian_1}
\end{equation}
The second to last term in the above two-line single equation represents the $n$ paths to surface
relaxation with probability $p_{S}$, and $p_{V}$ is the probability
per time for volume relaxation. Multiplying Eq.~(\ref{eq:BC_discrete_derivation_Cartesian_1})
with $\Delta r$, and then neglecting all but the leading terms gives
\begin{equation}
0=\sum_{j=D-n+1}^{D}\left[\frac{\Delta r^{2}}{2D\Delta t}\frac{M\left(\mathbf{x}-\Delta r \mathbf{e}_{j},t\right)-M\left(\mathbf{x},t\right)}{\Delta r}-\frac{\Delta r}{\Delta t}\frac{p_{S}}{2D}M\left(\mathbf{x},t\right)\right].
\label{e.long}
\end{equation}
\end{widetext}
Taking the limits $\Delta r\rightarrow 0,\:\Delta t\rightarrow 0$ (and $p_{S}\rightarrow 0,\: p_{V}\rightarrow 0$),
with $\Delta t/p_{V}=T_{V}$ and $\Delta r^{2}/\left(2D\Delta t\right)=D_{0}$ constant, we have above the Robin boundary condition
of Eq.~(\ref{eq:Diffusion_equation}) in each of the $n$ directions
$x_{j},\: j=D-n+1,...,D$, with the identifications $D_{0}=\Delta r^{2}/\left(2D\Delta t\right)$
and $\rho=\Delta rp_{S}/\left(2D\Delta t\right)$. 
Hence, combining these two relations we have
established the following surface relaxation relation for the
BC in each of the $n$ directions, 
\begin{equation}
p_{S}=\Delta r\rho/D_{0}. \label{p_S}
\end{equation} 
This leading order relation is not novel \cite{MendelsonPRB1990}, and
the higher order relation $\tilde{p_{S}}=\Delta r\rho/\left(D_{0}+\Delta r\rho\right)\simeq\sum_{j \geq1}\left(-1\right)^{j+1}\left(\Delta r\rho/D_{0}\right)^{j}$
have been suggested \cite{BanavarPRL1987}. 
For $\Delta r\rho\ll1$ the two
relations $p_{S}$ and $\tilde{p_{S}}$ are practically equivalent, but the latter perform slightly better
when benchmarked against the analytic result of Eq.~(\ref{eq:Solution_For_D_Cube})
for large $\rho$ (we used $\Delta r= 10^{-3} R_0$ for these tests). 

Finally we note that several researchers
apply an additional ``factor $3/2$'' in the surface relaxation relation (\ref{p_S}) for arbitrary digital domains, as was derived in \cite{BergmannPRE1995} in $D=3$ for continuous random walks.

\section{Local boundary conditions for digital domains}  \label{LLBC_4}

For random walk simulations to converge with high accuracy,
the number of trajectories needs to be large, and the step-size
($\Delta r$) needs to be small, see \cite{FinjordTPM2007,Feller1967}
for details. However, as we illustrate here for digitized media, the
way the true geometry is mapped onto for example a computed tomography (CT) digital image is of additional
importance. 
This is an intrinsic uncertainty, even if errors due to
segmentation \cite{Mueter2012} and resolution are neglected. 

We now
introduce a \emph{correction factor} $g \left(\mathbf{x}\right)$, to
be multiplied with the right hand side of the surface relaxation relation in Eq.~(\ref{p_S}), for an improved local Robin boundary
condition. With local we mean a local interpolated
surface, even though for example $\rho$ can locally also depend on space and time. 
Clearly $g \equiv 1$ corresponds to no correction, while an exact local value of $g \left(\mathbf{x}\right)$ requires that we know the exact surface locally. 
This is not the case for a general digital image from an application, but we here use two basic domains to evaluate the presented method.
In particular we have implemented a linear local boundary
condition (LLBC) for the discrete Cartesian random walk for which a linear interpolation of a general digital surface is implicit but no knowledge about the true surface is required.
More sophisticated correction factors, i.e., non-linear local boundary conditions corresponding to higher order interpolated surfaces, are possible. 
They may be motivated in future simulations if the input physical parameters are well known and high accuracy is required.
However, we have concluded by the comparisons with the analytic solutions for the basic domains that already the LLBC correct a substantial part of the error in the dynamics caused by the digital misrepresentation. 

As one way to evaluate the introduced local correction factor $g \left(\mathbf{x}\right)$ we introduce,
motivated by Eqs.~(\ref{eq:Asymptote_For_D_Ball}), (\ref{eq:Asymptote_For_D_Cube}) and (\ref{eq:Asymptote_General}), an initial
slope which also depends on the orientation of the Cartesian coordinate system
\begin{equation}
\frac{1}{\mathcal{M}\left(0\right)}\left.\frac{d\mathcal{M}\left(t,\phi_{1},...,\phi_{D(D-1)/2}\right)}{dt}\right|_{t=0}\equiv - f \rho\frac{S}{V},\label{eq:Angle_Dependent_Initial_Slope}
\end{equation}
where $f \left(\phi_{1},...,\phi_{D(D-1)/2}\right)$ is
defined above as a dimensionless global error factor dependent on the Euler angles in $D$-dimensions.
The error factor $f$ can quantify the error in the initial slope of the total magnetization caused by the digital misrepresentation for a uniform initial condition of the magnetic moment.
Note that the error factor $f$ is generally related to the correction factor $g \left(\mathbf{x}\right)$
in a non-trivial way.

\section{Results for linear local boundary conditions}  \label{RFTDD_5}

We present results only for $D=2$
here, while LLBC in higher dimensions is considered and numerically applied to true CT-images of porous media in an ongoing research project. 

\begin{figure}
\centerline{\includegraphics[scale=0.52]{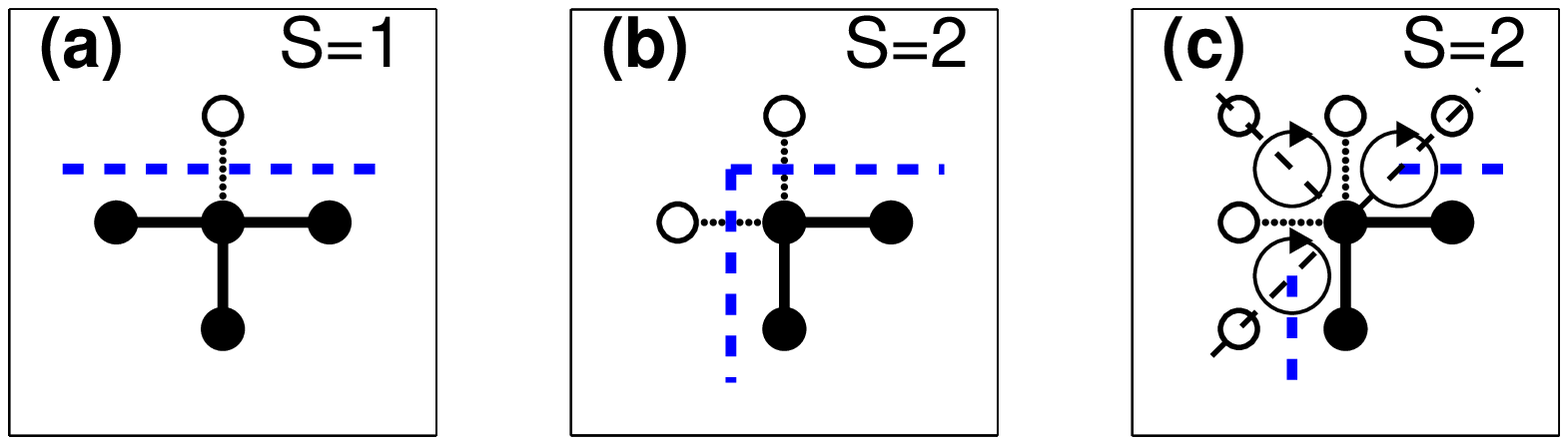}}\vspace{2mm}
\centerline{\hspace{0.3mm}~\includegraphics[scale=0.52]{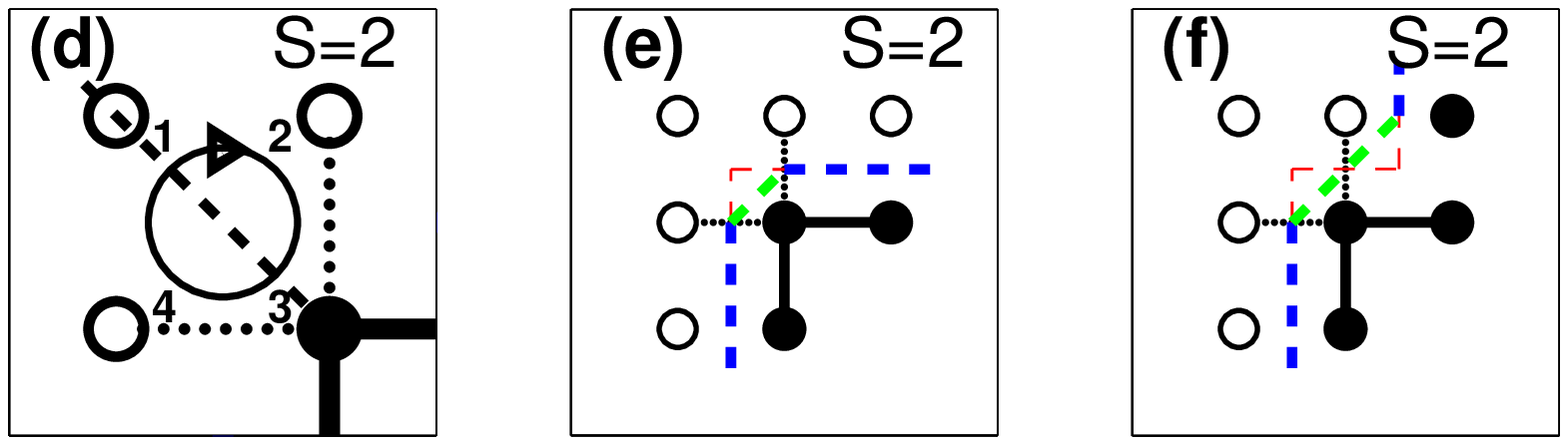}}
\caption{(Color online) Illustration of some pore-matrix interfaces for $D=2$. (a)~One of the
$4$ configurations for the number of neighboring surface (dashed lines) being $S=1$. (b)~One
of the $6$ configurations for $S=2$. (c)~Neighboring cells needed to
construct the LLBC for the case in (b). (d)~A LLBC-diagram illustrating the four lattice
points ``1'', ``2'', ``3'', ``4'' (see text) in one of the cells. 
In (e), corresponding to (b), and in (f)
we show examples of two different interpolated surfaces for $S=2$, e.g., in (f),
$p_{S}$ for a step upwards (to the left) is reduced by a factor $1/\sqrt{2}\simeq0.707$
($\sqrt{2}/4+1/2\simeq0.854$). \label{fig:Illustrations-of-some-possible-pore-matrix-interfaces}}
\end{figure}

In order to construct LLBC around a point $\mathbf{x}$ we need to distinguish between the $2^{2^D}$ possible lattice configurations in each cell surrounding $\mathbf{x}$ in each diagonal direction that is in contact with a boundary, see Figs.~\ref{fig:Illustrations-of-some-possible-pore-matrix-interfaces}(c) and (d). 
For this purpose we can locally define the integer $I\left(\mathbf{x}\right)=\sum_{j=1}^{2^D}Z_{j}2^{\left(2^D-j\right)}$, which for $D=2$ represents the $16$ different configurations $I\left(\mathbf{x}\right)\in\left\{ 0,1,...,15\right\} $,
where $Z_{j}$ is chosen to be $1$ ($0$) for a lattice point inside (outside) the pore volume, i.e., in Fig.~\ref{fig:Illustrations-of-some-possible-pore-matrix-interfaces}(d) for a black (white) dot around the point $\mathbf{x}$.
If $I\left(\mathbf{x}\right)\in\left\{ 1,2,4,7,8,11,13,14\right\} $,
it means that for any of these $8$ configurations we are locally going to interpolate a corner, see Figs.~\ref{fig:Illustrations-of-some-possible-pore-matrix-interfaces}(b)
and (e), for which we define the linear correction factor $g \left(\mathbf{x}\right)=1/\sqrt{2}$
(else $1$). Note that only half of the upper-left cell surface belongs
to the move-up boundary, whereas half belongs to the move-left boundary, consider the diagonal lines in Figs.~\ref{fig:Illustrations-of-some-possible-pore-matrix-interfaces}(c)
and~(d). 
The procedure we outline for locally generating these improved
linear boundaries is equivalent to what is known in $D=3$ as \emph{Marching
cubes}, generalised to arbitrary dimensions in \cite{Bhaniramka2004}. 

\begin{figure}
\includegraphics[scale=0.31]{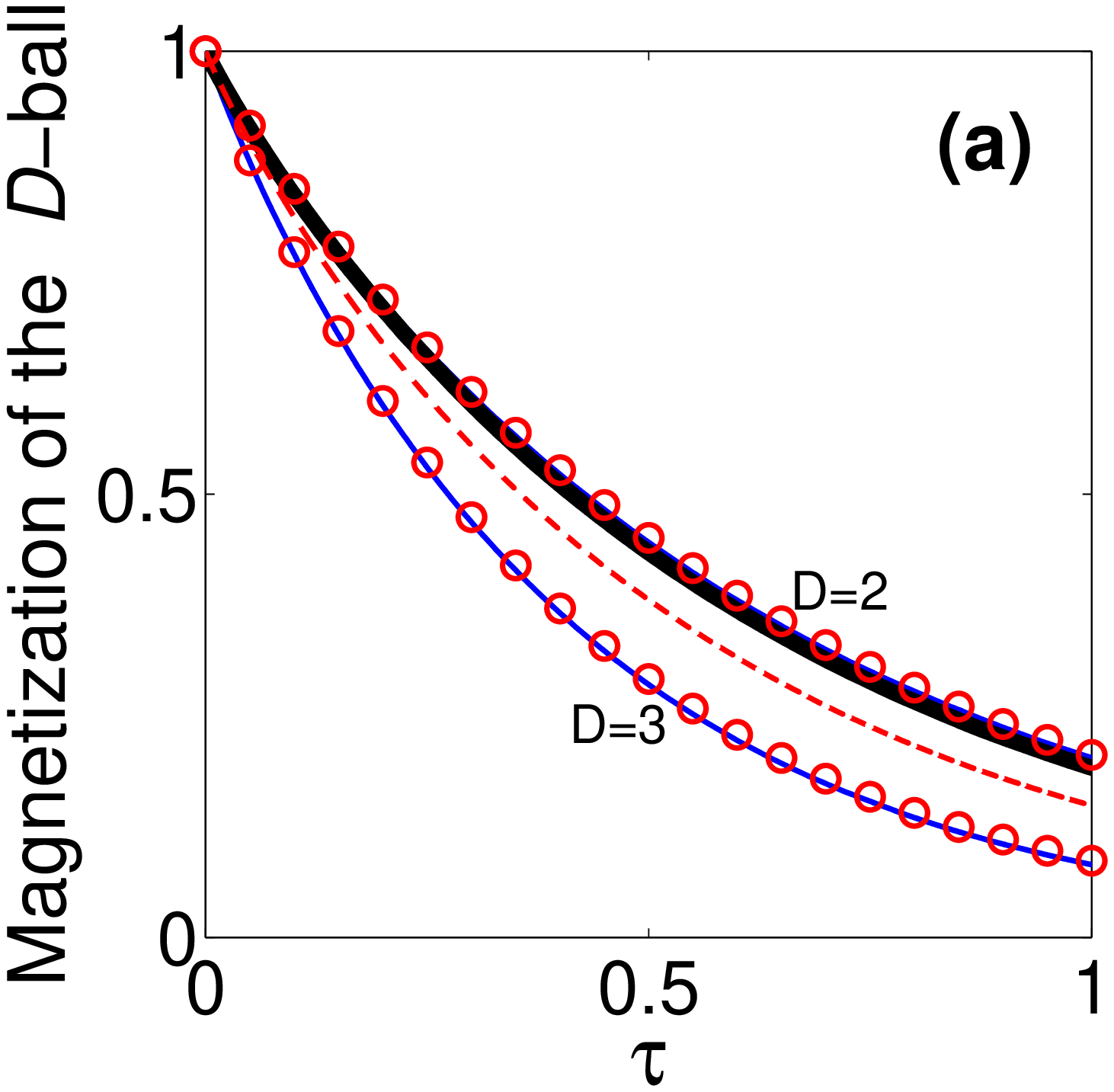}~\hspace{-3mm}~\includegraphics[scale=0.31]{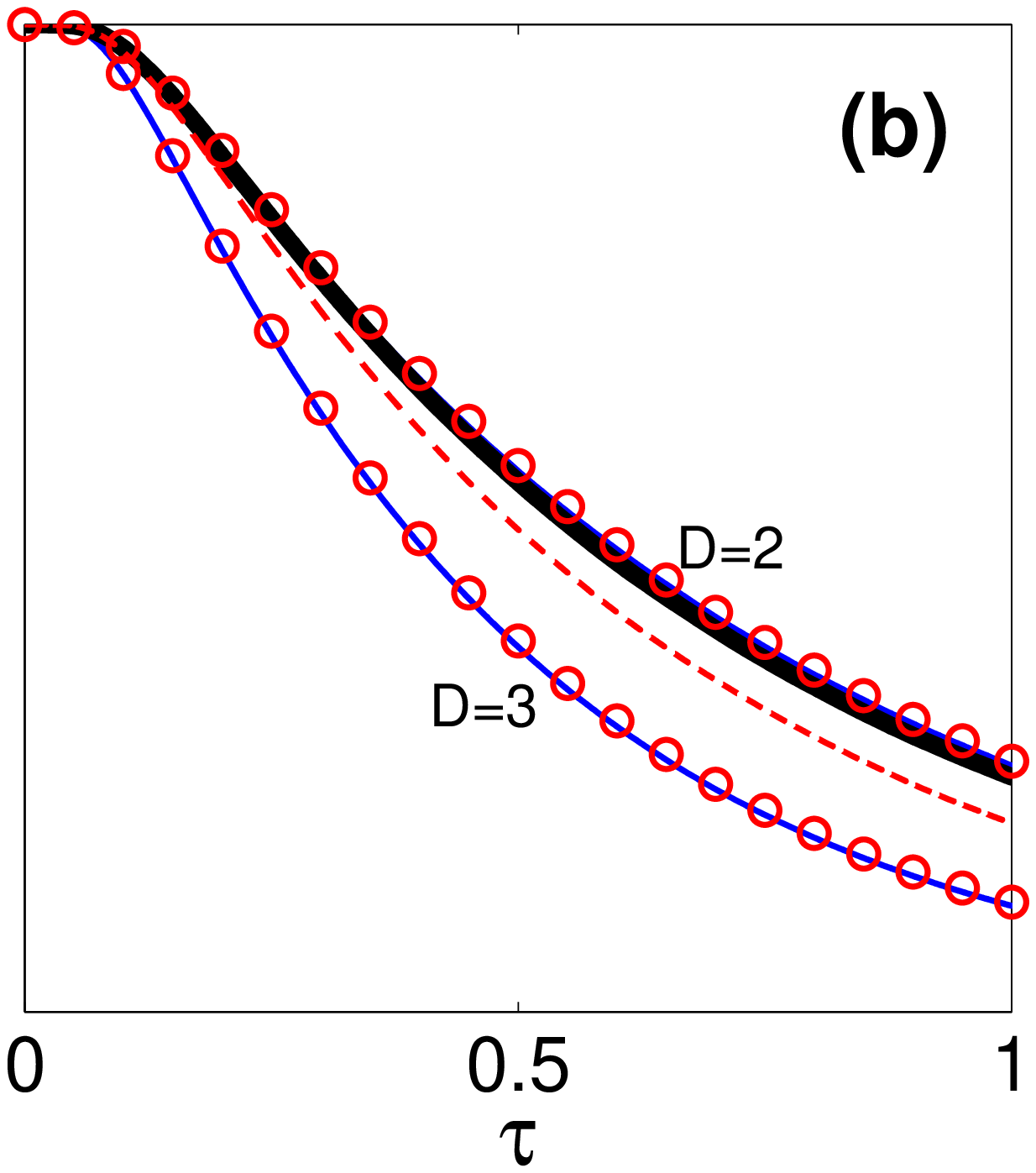}
\caption{(Color online) Dynamics of $\mathcal{M}\left(t\right)$ for different initial conditions as in Fig.~\ref{fig_1}.
Solid (blue) curves still shows results from the $D$-dimensional radial
random walk for reference. Open (red) rings shows results from the Cartesian random
walk in a digitized circle/sphere with $g =1/f$
(see text). 
For the circle ($D=2$) the results of using LLBC (see text) is shown with thick (black) curves slightly
below the $g =1/f_{c}$ results for $D=2$. Further below the thick LLBC curve is the uncorrected results ($g\equiv 1$) for the digitized circle, dashed (red) curves.  \label{fig_3}}
\end{figure}

The two basic domains with the random walks presented in sections~\ref{RRW_2} and~\ref{CRW_3} respectively, are trivial in the
sense that there are no local variations for the ratio of the true
pore surface and the digitized surface (except the corners of the
square in the Cartesian case). 
For simplicity, we then use $g \left(\mathbf{x}\right) = g =1/f $, i.e.,
with no space dependence, for a first numerical evaluation of the Cartesian random walk applied to the $D$-ball (for $D=2,3$), see Fig.~\ref{fig_3}. 

\begin{figure}
\includegraphics[scale=0.30]{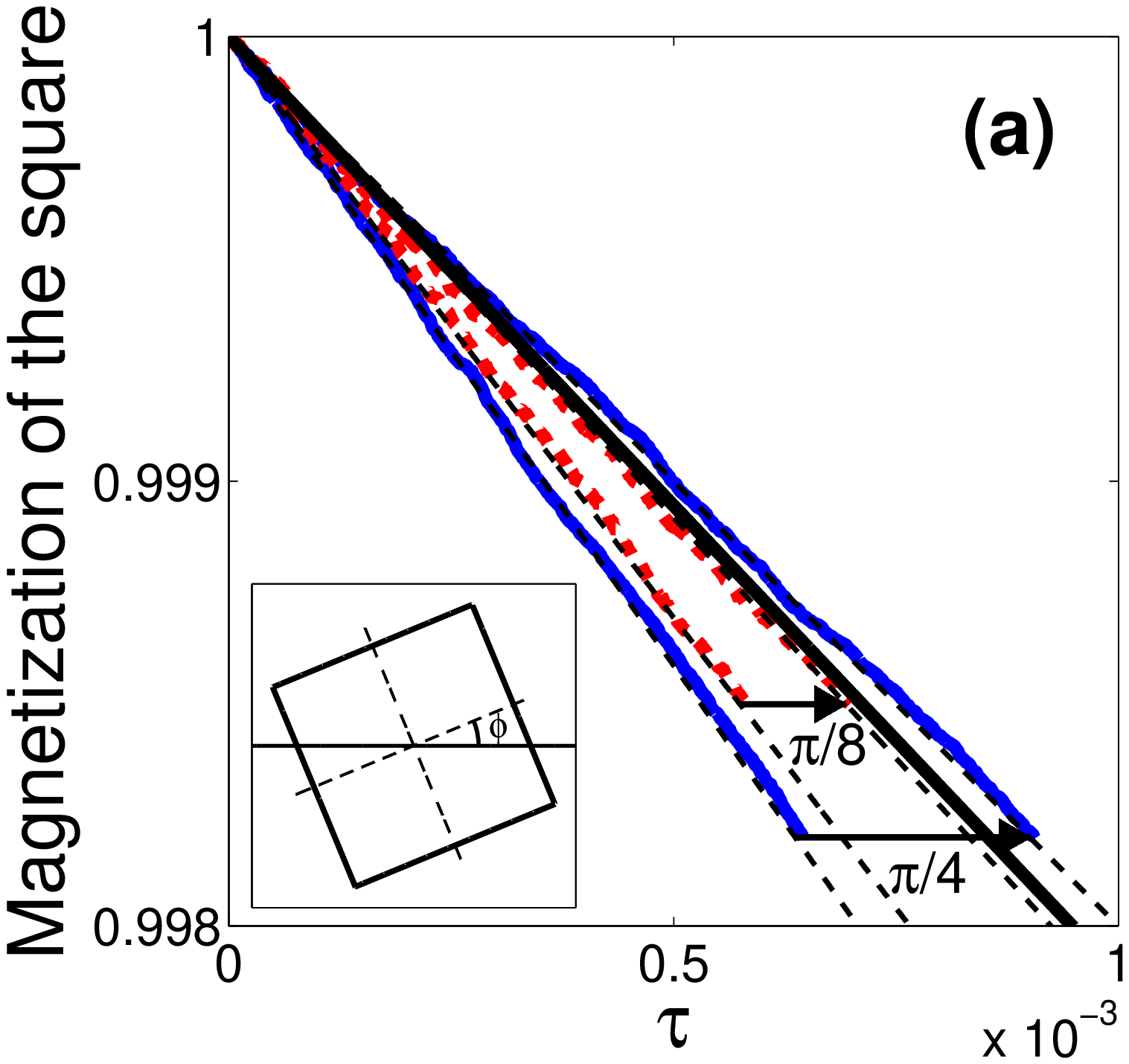}~\hspace{-2mm}~\includegraphics[scale=0.30]{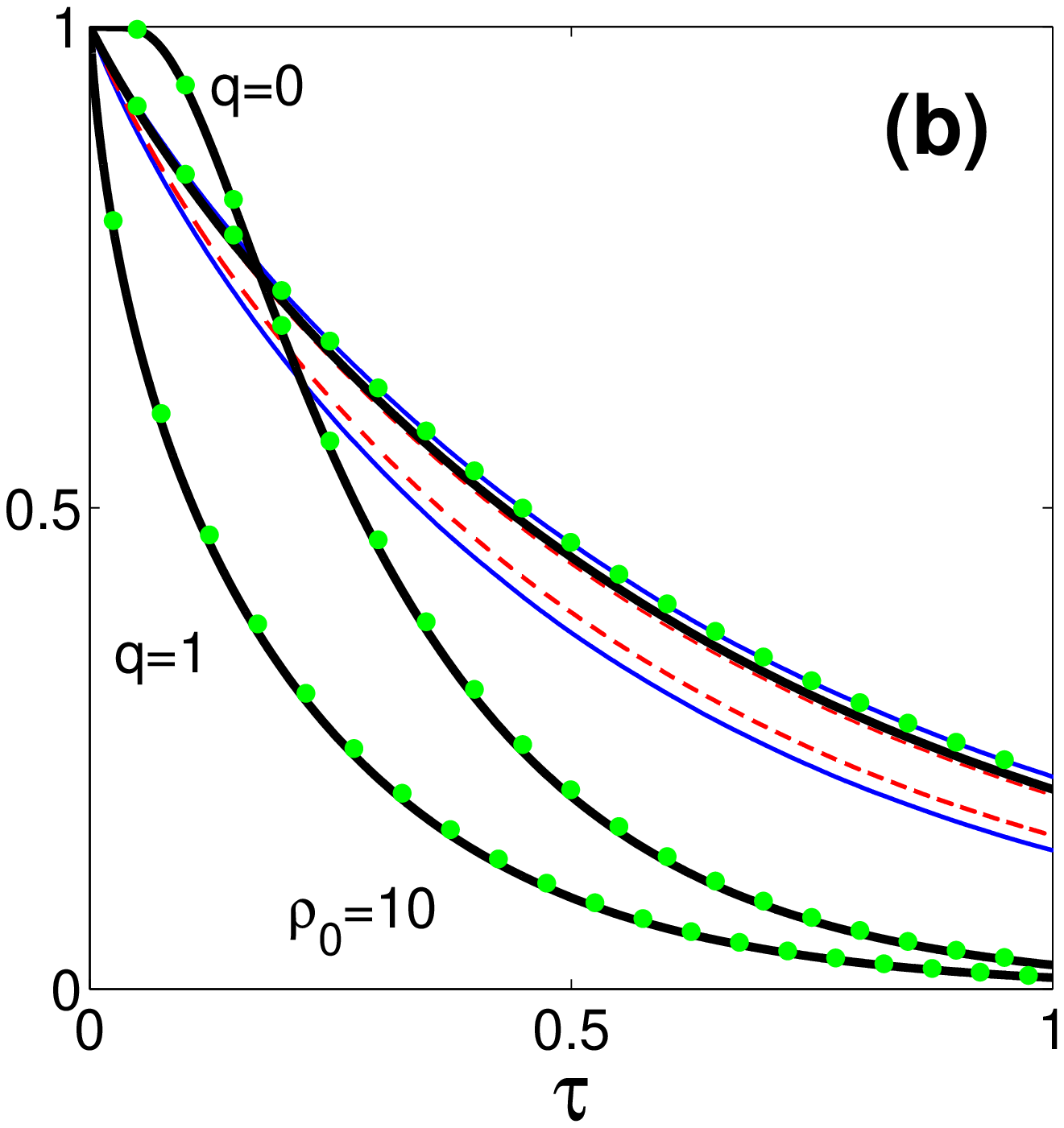}
\caption{(Color online) LLBC improvements for a square with different orientations
$\phi$, defined in the lower-left inset, are shown for small times in (a). 
Outer solid (blue) curves show numerical results for $\phi=\pi/4$, inner
dashed (red) curves are for $\phi=\pi/8$. In both cases the most left
curves are without LLBC ($g \equiv 1$), and the improvements by using LLBC
are indicated with arrows pointing on the corrected curves. Thin dashed
(black) lines shows analytic short-time asymptotes of Eq.~(\ref{eq:Angle_Dependent_Initial_Slope})
with $\rho S/V=2$ and (from left to right) the original error factors $f_{s} \left(\pi/4\right)=\sqrt{2}$,
$f_{s} \left(\pi/8\right)\simeq1.31$; respectively the error factors corresponding to LLBC: $f_{s} \left(\pi/8\right)\simeq1.08$,
$f_{s} \left(\pi/4\right)=1$, see text. 
The thick black curve is for an angle averaged LLBC with $\Delta\phi=\pi/160$.
The corresponding full timeinterval to (a) is shown in (b) by the upper right family of curves, and with (green) dots for the analytic
solution of Eq.~(\ref{eq:Solution_For_D_Cube}). 
For the additional two $\rho_{0}=10$ (slow diffusion) lower curves 
only analytic solutions (green dots) and angle averaged LLBC (thick black) results are shown. Here the curve near $q=1$ is for $m\left(\mathbf{x},0\right)=1/4$ while
$q=0$ corresponds to $m\left(\mathbf{x},0\right)=R_{0}^{2}\delta\left(\mathbf{x}\right)$.} \label{fig:_fig_3_square}
\end{figure}

For the remainder we concentrate the presentation on $D=2$. 
For the digitized
circle ($c$) the surface is $8R_{0}$ (same for the square) and hence
the digitized-to-true surface ratio gives the error factor $f_{c} =4/\pi\simeq1.27$.
After geometric considerations we obtained the corresponding LLBC ratio for the circle
 to be $f_{c} =8\left(\sqrt{2}-1\right)/\pi\simeq1.05$.
For the square ($s$), we illustrate the single angle variable $\phi$, i.e., $D(D-1)/2=1$ here, in the lower-left inset of Fig.~\ref{fig:_fig_3_square}. 
One can then show with geometry that the digitized-to-true surface ratio follows
the $\pi/2$-periodic error factor $f_{s} \left(\phi\right)=\sqrt{2}\cos\left(\phi-\pi/4\right),\: 0\leq\phi<\pi/2,\: f_{s} \left(\phi+\pi/2\right)=f_{s} \left(\phi\right)$,
For the LLBC ratio one instead obtain the $\pi/4$-periodic error factor $f_{s} \left(\phi\right)=\sqrt{2}\left[\cos\left(\phi+\pi/4\right)+\sin\left(\phi\right)\right]$,
with $\max( f_{s} )=f_{s} \left(\pi/8\right)\simeq1.08$.

To further minimize the risk of retrieving the worst case scenario
in a simulation of an unknown digitized medium one can in principle examine all different
orientations by varying the $D(D-1)/2$ Euler angles of the coordinate system. For the digitized square, a uniform average over the single Euler angle $\phi$
still overestimates the surface with $\langle f_{s} \rangle_{\phi}=f_{c} \simeq1.27$,
but only with $\langle f_{s} \rangle_{\phi}=f_{c} \simeq1.05$
using LLBC. 

The above results are strictly valid in the $\Delta r\rightarrow0$
limit but showed good agreement with the numerical results that are presented in Fig.~\ref{fig:_fig_3_square} for $\Delta r=10^{-2}R_{0}$.

\section{Discussion and summary}  \label{DAS_6}
While the LLBC can
always be used, the exact correction factor can only be used if the true surface is known locally. However,
for a complex porous medium where the variations of $\rho\left(\mathbf{x}\right)$
may also be partly unknown, a measurement of the total pore
surface can then be useful. The initial slope from a NMR relaxation measurement
probes a combined effect of $\rho\left(\mathbf{x}\right)$ and $S/V$, see Eq.~(\ref{eq:Angle_Dependent_Initial_Slope}), so measuring
the surface-to-volume ratio can alone provide a first improvement.
However, when comparing with measurements, keep in mind that a molecule carrying magnetic spin has a certain
lengthscale (crossection) for surface relaxation, whereas for example common BET
techniques \cite{Brunauer1938} maps out the surface dependent on
the size of the molecule in use (e.g. $\sim0.2$ nm for $N_{2}$).

Even when a correct (algorithm dependent) relation between
$\rho$ and $p_{S}$ is used, the accuracy in diffusion simulations
with Robin boundary conditions is severely restricted. This is due
to an uncertainty of the true $\left(D-1\right)$-surfaces in a $D$-dimensional
digitized media. As illustrated with the basic domains, the accuracy can
be increased substantially by introducing a linear local correction
to the relaxation on a digital surface. An improved interplay between
experimental measurements and higher order algorithms for local Robin boundary conditions will increase
the accuracy of simulations, applied to surface reactions and NMR
dynamics in porous media and MRI-based analysis in medicine, beyond the first step taken here. 

Finally we remark that local effects appears more dramatic if one for example directly study functions of $M\left(\mathbf{x},t\right)$ instead of its spatial
integral.

\section*{Acknowledgement}

We are grateful for funding from P$^3$---\textit{Predicting Petrophysical Parameters}, supported by the Danish Advanced Technology Foundation (HTF) and Maersk Oil.
We thank M. Gulliksson and an anonymous referee for valuable comments.

\end{document}